\documentclass[pra,showpacs,amsfonts,amsmath,floatfix]{revtex4}

\usepackage{graphicx}
\usepackage{color}

\begin{document}

%=======================================================================================
\title{Reflection of a Lieb-Liniger wave packet from the hard-wall potential}

\author{D.~Juki\'{c}}
\affiliation{Department of Physics, University of Zagreb, Bijeni\v cka c. 32, 10000 Zagreb, Croatia}

\author{H.~Buljan}
\email{hbuljan@phy.hr}
\affiliation{Department of Physics, University of Zagreb, Bijeni\v cka c. 32, 10000 Zagreb, Croatia}

\date{\today}

\begin{abstract}
Nonequilibrium dynamics of a Lieb-Liniger system in the presence of the 
hard-wall potential is studied. 
We demonstrate that a time-dependent wave function, which 
describes quantum dynamics of a Lieb-Liniger wave packet comprised of $N$ particles, 
can be found by solving an $N$-dimensional Fourier transform; 
this follows from the symmetry properties of the many-body eigenstates 
in the presence of the hard-wall potential. 
The presented formalism is employed to numerically calculate reflection of a few-body 
wave packet from the hard wall for various interaction strengths and incident momenta. 
\end{abstract}

\maketitle
%\narrowtext
%\newpage

%=======================================================================================
\section{Introduction}
\label{sec:intro}

Recent intensive theoretical studies of one-dimensional (1D) Bose gases 
have been motivated mainly by the experimental realizations 
\cite{OneD,TG2004,Kinoshita2006,Hofferberth2007,Amerongen2008} 
of the models \cite{Lieb1963,Girardeau1960} describing these systems. 
Ultracold atomic gases in tight atomic wave guides can have their 
transversal degrees of freedom essentially frozen such that their motion becomes 
effectively one-dimensional \cite{OneD,TG2004,Kinoshita2006,Hofferberth2007,Amerongen2008}. 
These systems are described in terms of the Lieb-Liniger model, i.e., a system of identical Bose particles 
in 1D with pointlike $\delta$-function interactions of arbitrary strength $c$ 
\cite{Lieb1963}. In the limit of strong coupling $(c\rightarrow \infty)$, the 
Lieb-Liniger gas approaches the Tonks-Girardeau regime of impenetrable 
bosons \cite{Girardeau1960}, which has been experimentally achieved as well 
\cite{TG2004}; the Tonks-Girardeau regime occurs at very low temperatures, low linear densities, 
and with strong effective interactions \cite{Olshanii,Petrov,Dunjko}. 
Experiments are even capable of exploring nonequilibrium quantum dynamics of 
these 1D many-body systems \cite{Kinoshita2006,Hofferberth2007}, which may occur 
after some sudden change in the system's parameters. These ultracold atomic assemblies are 
well isolated from the environment, that is, their quantum coherence stays 
preserved for long times. Therefore, they may serve as a playground to investigate 
relaxation of isolated quantum many-body systems, which is one of the most 
interesting questions in theoretical physics (e.g. see Refs. \cite{RigolNature,
Igloi2000,Sengupta,Rigol2006,Cazallila2006,Calabrese2006,Cherng2006,Kollath2007} 
and references therein). 
Subsequent relaxation of 1D gases via collisions is greatly determined by the 
reduced dimensionality and the integrability of the underlying models. 
We are motivated to study the time-dependent Lieb-Liniger model because 
(i) today's experiments can explore fundamental physical questions in these 
systems \cite{Kinoshita2006,Hofferberth2007}, and (ii) one can construct exact 
solutions of some relevant problems for all interaction strengths (from the 
mean field regime up to the strongly correlated regime) \cite{Gaudin1983,
Girardeau2003,Buljan2008,Jukic2008,Jukic2009,Gritsev2009}. 

The eigenstates of the Lieb-Liniger model (without an external potential present), 
which were constructed by employing the Bethe ansatz \cite{Lieb1963}, 
are determined by a set of quasimomenta; when periodic \cite{Lieb1963} 
boundary conditions are imposed, the quasimomenta must obey a set of 
transcendental Bethe equations \cite{Lieb1963}. 
The Lieb-Liniger eigenstates in the presence of the hard-wall (i.e., 
on the semi-infinite line) can be constructed via superposition of 
free space eigenstates \cite{Gaudin1971}; again, if the quasimomenta 
should obey a particular set of transcendental equations \cite{Gaudin1971}, 
this superposition yields eigenstates in an infinitely deep box \cite{Gaudin1971}. 
Recent years have witnessed an increasing interest in exact solutions of these models 
(e.g., see \cite{Busch1998,Muga1998,Sakmann2005,Batchelor2005,Kinezi2006,
Sykes2007,Kanamoto2009} and references therein), most of which are focused on 
the properties of the ground and excited eigenstates
(see also Refs. \cite{Korepin1993,Gaudin1971a}). 
Unfortunately, the Lieb-Liniger model does not reveal exact solutions in the presence of 
some external trapping potential $V(x)$ (e.g., the harmonic potential).

In the Tonks-Girardeau limit $c\rightarrow\infty$, the methods for finding 
eigenstates \cite{Girardeau1960}, time-dependent solutions \cite{Girardeau2000}, 
as well as observables (e.g., see Ref. \cite{Rigol2005} for the system of 
hard-core bosons on the lattice and \cite{Pezer2007} for the continuous Tonks-Girardeau model) 
are much simpler due to the Fermi-Bose mapping, which in a simple fashion maps 
a fermionic wave function describing spinless noninteracting fermions onto a 
Tonks-Girardeau wave function \cite{Girardeau1960,Girardeau2000}. 
It is important to emphasize that these methods are valid for any external potential. 
Perhaps the simplicity of the methods and phenomenological relevance of the model 
\cite{Kinoshita2006} have lead to increasing interest in quantum many-body dynamics 
of Tonks-Girardeau gases. Some of these studies include dynamics during free expansion 
\cite{Rigol2005,Minguzzi2005,DelCampo2006,Gangardt2008}, dynamics of dark soliton-like states 
\cite{Girardeau2000}, and reflections from a periodic potential \cite{Pezer2007}.

In the case of finite interaction strength $c$, it is far more difficult to calculate 
exact many-body wave functions and/or observables describing dynamics of 
time-dependent Lieb-Liniger wave packets. Without attempting to provide a review, let us 
mention a few approaches utilized to study nonequlibrium dynamics 
of 1D interacting Bose gases. The hydrodynamic formalism \cite{Dunjko} (the local 
density approximation) can be formulated in terms of the Nonlinear Schr\" odinger 
like equation with variable nonlinearity \cite{Ohberg2002}; this approach reduces 
to the Gross-Pitaevskii theory in the weakly interacting limit \cite{Dunjko,Ohberg2002}. 
More sophisticated numerical approaches include the time-evolving block decimation 
algorithm \cite{Vidal2004}, which has recently been utilized to study relaxation 
following a quench in a 1D Bose gas \cite{Muth2009}, the two–particle irreducible 
(2PI) effective action approach \cite{AMRey2004,Gasenzer2005}, 
the multiconfigurational time-dependent Hartree method for bosons (MCTDHB) 
\cite{Alon2008} (the MCTDHB method is numerically exact when sufficiently 
many time-dependent orbitals are taken into account), and 
the multiconfigurational time-dependent Hartree method (e.g., see 
Ref. \cite{Zollner2008} and references therein). 
Reference \cite{Gritsev2009} provides a discussion of several methods 
which can be used to describe nonequilibrium dynamics of Lieb-Liniger gases 
with greater focus on the form-factor approach \cite{Gritsev2009}, 
which has been recently utilized to calculate equilibrium correlation functions
of a 1D Bose gas (see \cite{Caux2007} and references therein). 
A broader review discussing many-body physics with ultracold gases can 
be found in Ref. \cite{Bloch2008}. We also mention a recent review on quantum 
transients \cite{delCampo2009}. 

An interesting exact method has been outlined by Gaudin way back in 1983 \cite{Gaudin1983}:
A time-dependent Lieb-Liniger wave function on an infinite line, in the absence 
of an external potential, can be constructed by acting with a differential 
operator (which contains the interaction strength parameter $c$) onto 
a time-dependent wave function describing noninteracting (spin polarized) 
1D fermions \cite{Gaudin1983,Buljan2008,Jukic2008,Jukic2009,Pezer2009}. 
For dynamics of a Lieb-Liniger wave packet comprised of $N$ particles, this method reduces 
to finding an $N$-dimensional Fourier transform, which can be used to extract 
the asymptotic behavior of the wave function and some observables 
during the course of 1D free expansion \cite{Jukic2008,Jukic2009}. 
In this article we investigate the possibility of extending this approach 
to study dynamics of a Lieb-Liniger wave packet in the presence of the 
hard-wall potential. 

Our interest in quantum dynamics in the presence of the hard-wall potential is 
in part motivated by experiments.
More specifically, the interaction of Bose-Einstein condensates (BEC) with surfaces 
is of interest for implementations of atom interferometry on chips \cite{Wang2005}. 
A BEC falling under gravity, and then reflecting from a light-sheet, has been 
experimentally and theoretically studied in Ref. \cite{Bongs1999}. 
Moreover, one of the prominent experimental activities nowadays is deceleration 
of atomic beams by reflection from a moving mirror. This work first started 
with neutrons being cooled by reflecting from a moving Ni surface \cite{neutron}.
In cold atoms physics, there have been several experiments for manipulation 
and slowing down atomic beams with the use of reflection mirrors 
\cite{Libson2006,Narevicius2007,Reinaudi2006}.

Here we explore, by using exact methods, dynamics of Lieb-Liniger wave packets 
in the presence of the hard-wall potential, more specifically, reflection of a Lieb-Liniger 
wave packet from such a wall. The outline of the paper is as follows. 
In Sec. \ref{sec:LLmodel} we introduce the model and outline the construction 
of eigenstates in the given external potential.
In Sec. \ref{sec:timedyn} we analytically discuss time-dependent quantum dynamics 
of the system which starts from a general initial condition.
By employing the symmetries of the Lieb-Liniger eigenstates, we demonstrate that 
a time-dependent Lieb-Liniger wave packet reflecting from the wall can be 
calculated by solving an $N$-dimensional Fourier transform, where $N$ is 
the number of particles. This opens the way to calculate the 
asymptotic properties of the wave packet by employing the stationary phase
approximation as in Refs. \cite{Jukic2008,Jukic2009} for free expansion. 
In Sec. \ref{sec:example} we utilize the formalism to numerically study 
dynamics of single-particle density and momentum distribution of a 
few-body wave packet reflecting from the wall. We find that 
the wave packets for smaller interaction strength $c$ get reflected at a 
slower rate, because they get compressed more strongly as the wave packet 
hits the wall. The interference fringes which occur during the dynamics 
have larger visibility for smaller values of $c$.

%=======================================================================================
\section{Eigenstates in the presence of the hard-wall potential}
\label{sec:LLmodel}
The Lieb-Liniger model describes $N$ identical bosons in one spatial dimension, 
which interact via a repulsive $\delta$-function potential of strength $c>0$.
The model can be represented in terms of the Schr\" odinger equation:
\begin{equation}
i \frac{\partial \psi}{\partial t}=
- \sum_{i=1}^{N}\frac{\partial^2 \psi}{\partial x_i^2} 
+ \sum_{1\leq i < j \leq N} 2c\,\delta(x_i-x_j)\psi 
+ \sum_{i=1}^N V(x_i) \psi(x_1,\ldots,x_N,t).
\label{LLmodel}
\end{equation}
As we have already stated, this model can be experimentally realized 
with ultracold atoms in tight atomic waveguides. The spatial and temporal 
coordinates ($x$ and $t$, respectively), as well as the potential $V(x)$ are 
dimensionless in this paper. Their connection to physical units is as follows: 
$x=X/X_0$, $t=T/T_0$, and $V(x)=U(X)/E_0$, where $X$, $T$ and $U(X)$ are space, time, 
and energy variables in physical units. Given the mass of the atoms $m$, 
the choice of an arbitrary length scale $X_0$ sets the time scale $T_0=2mX_0^2/\hbar$, 
and energy scale $E_0=\hbar^2/(2mX_0^2)$. 
Suppose that the transverse confinement of the atomic waveguide is 
described by a harmonic oscillator with frequency $\omega_{\perp}$. 
The interaction parameter $c$ is proportional to the effective 1D coupling strength $g_{1D}$ \cite{Olshanii},
$2c=g_{1D}/(X_0 E_0)=g_{1D} \ 2mX_0/\hbar^2$, which is related to the 1D scattering 
length $a_{1D}$ via $g_{1D}=-2\hbar^2/m a_{1D}$; the 1D scattering length 
$a_{1D}=-(l_{\perp}^2/a) (1-Ca/\sqrt{2}l_{\perp})$ depends on three-dimensional 
scattering length $a$ and the transverse oscillator width $l_{\perp}=\sqrt{\hbar/m \omega_{\perp}}$ 
(the constant $C=1.4603\ldots$).

In the present paper, we focus ourselves on the dynamics (in time) of a  
Lieb-Liniger wave packet in the presence of the hard-wall potential, that is, 
\begin{align}
V(x)= 
\begin{cases} 
0, & \mbox{if } x>0 \\ 
\infty, & \mbox{if } x \leq 0. 
\end{cases} 
\label{potential}
\end{align}
We will show that the solution of this problem can be constructed by 
solving an $N$-dimensional Fourier transform. 
To this end, we need eigenstates of a Lieb-Liniger gas in the hard-wall 
potential. First, let us write down the Lieb-Liniger eigenstates 
in free space (i.e., $x\in(-\infty,\infty)$ without external potentials and 
any boundary conditions):
\begin{eqnarray}
\psi_{\{ k \}} & = & \mathcal{N}(\{k\}) \sum_P (-1)^P
\prod_{i<j}{\left[\mbox{sgn}(x_j-x_i)+\frac{i}{c} (k_{Pj}-k_{Pi}) \right]} 
e^{i \sum_j k_{Pj}x_j} \nonumber \\
 & = & \mathcal{N}(\{k\}) \sum_P (-1)^P a(P,\{ k \}) e^{i \sum_j k_{Pj}x_j}
\label{psi_k}
\end{eqnarray}
where $\{ k \}=\{ k_m\, |\, m=1,\ldots,N \}$ is a set of (real) distinct
quasimomenta which uniquely determine the eigenstate, 
$P$ denotes a permutation of $N$ numbers, $P\in S_N$, 
and we have implicitly defined $a(P,\{ k \})$.
The normalization of these eigenstates is given by \cite{Korepin1993,Gaudin1971a}
\[
\frac{1}{\mathcal{N}({\{ k \}})}=
\sqrt{ N! (2 \pi)^N \prod_{i<j}{\left[1+\left(\frac{k_j-k_i}{c}\right)^2 \right]}},
\] 
that is, within the fundamental sector in $k$-space,
$k_1<\cdots<k_N$ and $k'_1<\cdots<k'_N$, we have
\begin{align}
\int_{-\infty}^{\infty} \psi_{\{k\}}^{*} \psi_{\{k'\}}	
dx_1 \cdots d x_N = \prod_{j=1}^N{\delta(k_j-k'_j)}.
\label{norm_line}
\end{align}

The Lieb-Liniger eigenstates in the presence of the hard-wall (denoted by $\phi_{\{k\}}$) were 
first constructed by Gaudin \cite{Gaudin1971} as a superposition of $2^N$ 
free-space eigenstates. This superposition obeys the hard-wall boundary condition: 
$\phi_{\{k\}}(x_1=0,x_2,\ldots,x_N)=0$ in the fundamental sector 
$R_1:x_1<x_2<\ldots<x_N$ of $x$-space. 
These eigenstates are expressed as follows:
\begin{equation}
\phi_{\{k\}} = \sum_{\{ \epsilon \}} A(\{ \epsilon \},\{ k \}) \psi_{\{\epsilon k\}},
\label{psiHW}
\end{equation}
where $\{\epsilon \}=\{ \epsilon_m \, |\, \epsilon_m\in\{ -1,1 \}, m=1,\ldots,N \}$
and 
$\{\epsilon k\}=\{ \epsilon_m k_m \, |\, \epsilon_m\in\{ -1,1 \}, m=1,\ldots,N \}$; 
evidently, there are $2^N$ such sets and therefore $2^N$ terms in the sum (\ref{psiHW}). 
The quantity $A(\{ \epsilon \},\{ k \})$ is defined by 
\begin{equation}
A(\{ \epsilon \},\{ k \}) = \epsilon_1  \cdots \epsilon_N \ A'(\epsilon_1 k_1,\epsilon_2 k_2,\ldots,\epsilon_N k_N),
\label{Acrtano}
\end{equation}
where
\begin{equation}
A'(k_1,k_2,\ldots,k_N) \equiv \frac{\prod_{i<j}{\left[1+\frac{i}{c} (k_j+ k_i) \right]}}
	    {\sqrt{\prod_{i<j}{\left[1+\left(\frac{k_j+k_i}{c}\right)^2 \right]}}},
\end{equation}
are the coefficients utilized in the superposition. 
It is straightforward to verify that indeed $\phi_{\{k\}}(x_1=0,x_2,\ldots,x_N)=0$ in the 
fundamental sector $R_1$ \cite{Gaudin1971}. 

However, it is not simple to prove that these eigenstates are orthogonal and 
normalized. This is of key importance if one wishes to project 
some initial state onto these eigenstates and calculate time-evolution 
in the standard fashion via superposition over eigenstates. 
In Section \ref{sec:example} we discuss the normalization of eigenstates (\ref{psiHW}), 
and based on our numerical investigations conjecture that these 
eigenstates are orthogonal and normalized.

%=======================================================================================
\section{Many-body dynamics in time via a Fourier transform}
\label{sec:timedyn}

In this section we demonstrate that a solution of the time-dependent equation 
(\ref{LLmodel}) with the hard-wall potential (\ref{potential}) can be expressed 
in terms of an $N$-dimensional Fourier transform. 
We assume that at time $t=0$ the wave packet is localized in the vicinity of the wall. 
For example, the initial state $\psi_0$ can be the ground state wave function 
in some external trapping potential; if at $t=0$ this potential is suddenly 
turned off, the wave packet will start expanding and some of its components 
will be reflected from the wall which will give rise to interference effects. 
Such a scenario is possible to create with today's experimental 
capabilities \cite{Kinoshita2006}. One possible (similar) scenario is as 
follows: suppose that at $t=0$ the aforementioned trapping potential is turned off, 
and that in the next instance the many body wave packet is given some momentum 
kick, say towards the wall; the reflection and interference phenomena 
will depend on the interactions and imparted momentum. 
During the reflection, particles will collide and one may ask to which 
extent will the initial conditions be forgotten (or blurred) after the 
reflection?

To describe quantum dynamics from the initial conditions described above, we write the
initial state $\psi_0$ as a superposition over complete set of eigenstates $\phi_{\{k\}}$:
\begin{equation}
\psi_0 = \int_{0<k_1<\cdots<k_N} b(k_1,\ldots,k_N) \phi_{\{k\}} dk_1\ldots dk_N. 
\label{initB0}
\end{equation}
The subsequent derivation is based on the following two relations obeyed by 
the eigenstates $\phi_{\{k\}}$:
\begin{equation}
\phi_{\{k\}} =(-1)^P \phi_{\{P k\}},
\label{perm_G}
\end{equation}
and
\begin{equation}
\phi_{\{k\}} = -\phi_{\{ k_1,\ldots,k_{j-1},-k_j,k_{j+1}\ldots,k_N \}}.
\label{minus_G}
\end{equation}
Equation (\ref{perm_G}) follows from the definition of $\phi_{\{k\}}$ in 
Eq. (\ref{psiHW}), and the fact that the Lieb-Liniger eigenstates in free space 
$\psi_{\{k\}}$ obey identical relation: $\psi_{\{k\}} =(-1)^P \psi_{\{P k\}}$; 
this identity can be traced to the fact that $\psi_{\{k\}}$ are antisymmetric 
with respect to the interchange of any two variables $k_i$ and $k_j$ 
\cite{Korepin1993}. The derivation of Eq. (\ref{minus_G}) is straightforward. 
Let us define a set $\{ \epsilon' \}$, which corresponds to the set $\{\epsilon\}$ 
as follows: $\{ \epsilon'_1,\ldots,\epsilon'_N \}=
\{ \epsilon_1,\ldots,\epsilon_{j-1},-\epsilon_j,\epsilon_{j+1},\ldots,\epsilon_N \}$;
it is evident from the definition (\ref{psiHW}) that $\phi_{\{k \}}=
\sum_{\{ \epsilon' \}} A(\{ \epsilon' \},\{ k \}) \psi_{\{\epsilon' k\}}$. 
Furthermore, let us denote $\{ k' \}=\{ k_1,\ldots,k_{j-1},-k_j,k_{j+1}\ldots,k_N \}$, 
i.e., the set of $k$-values $\{ k' \}$ is identical to the set $\{ k \}$ except 
that $k_j$ is reversed in sign. 
By using $A(\{ \epsilon \},\{ k \})=-A(\{ \epsilon' \},\{ k' \})$ 
and $\{\epsilon' k'\}=\{\epsilon k\}$ we have 
\begin{equation}
\phi_{\{k' \}}  =  \sum_{\{ \epsilon' \}} A(\{ \epsilon' \},\{ k' \}) \psi_{\{\epsilon' k' \}} 
 =  - \sum_{\{ \epsilon \}} A(\{ \epsilon \},\{ k \}) \psi_{\{\epsilon k \}} =-\phi_{\{k \}},
\end{equation}
that is, we obtain Eq. (\ref{minus_G}). We note in passing that 
if any $k_j=0$, then $\phi_{\{k\}}=0$, which follows from Eq. (\ref{minus_G}); 
furthermore, $\phi_{\{k\}}$ is also zero whenever any two of the quasimomenta 
$k_i$ and $k_j$ are equal.

Due to the symmetry of the hard-wall eigenstates $\phi_{\{k\}}$ presented in 
Eqs. (\ref{perm_G}) and (\ref{minus_G}), a complete set of eigenstates 
is spanned in the region of the $k$-space defined by $0<k_1<\ldots<k_N$, which 
we will refer to as the fundamental region in $k$-space, and denote it with 
$Q_1^+$. Hence, the integral in Eq. (\ref{initB0}) spans over $Q_1^+$. 
Furthermore, by employing relations (\ref{perm_G}) and (\ref{minus_G}), $\psi_0$ can be 
written as an integral over the whole $k$-space:
\begin{equation}
\psi_0 = \int_{-\infty}^{\infty} dk_1 \cdots dk_N G(k_1,\ldots,k_N) e^{i \sum_j k_j x_j},
\label{psiB0-G}
\end{equation}
where the function $G$ is defined as
\begin{equation}
G(k_1,\ldots,k_N) = b(k_1,\ldots,k_N) A'(k_1,\ldots,k_N)
\mathcal{N}(\{k\}) \prod_{i<j}{\left[\mbox{sgn}(x_j-x_i)+\frac{i}{c} (k_j-k_i) \right]}.
\label{Gk}
\end{equation}
From Eq. (\ref{Gk}) it immediately follows that the time-evolution of a 
Lieb-Liniger wave packet in the presence of the hard-wall can be calculated from an 
$N$-dimensional Fourier transform. 

In order to derive Eqs. (\ref{psiB0-G}) and (\ref{Gk}), first note that 
due to (\ref{perm_G}) and (\ref{minus_G}), the projection coefficients 
satisfy
\begin{equation}
b(k_1,\ldots,k_N)=(-1)^P b(k_{P1},\ldots,k_{PN}),
\label{b_perm}
\end{equation}
and
\begin{equation}
b(k_1,\ldots,k_N) = - b(k_1,\ldots,k_{j-1},-k_j,k_{j+1},\ldots,k_N);
\label{b_minus}
\end{equation}
the latter identity can conveniently be rewritten as 
\begin{equation}
b(k_1,\ldots,k_N) = \epsilon_1  \cdots \epsilon_N 
                    b(\epsilon_1 k_1,\ldots,\epsilon_N k_N).
\label{b_eps}
\end{equation}

By employing the symmetries of the Lieb-Liniger hard-wall eigenstates, 
which are inherited by the expansion coefficients $b(k_1,\ldots,k_N)$,
Eq. (\ref{initB0}) can be rewritten as follows:
\begin{widetext}
\begin{align}
\psi_0 	&= \frac{1}{N!} \int_{k_1>0,\ldots,k_N>0} dk_1\cdots dk_N b(k_1,\ldots,k_N)\ \sum_{\{ \epsilon \}} A(\{ \epsilon \},\{ k \}) \nonumber  \\ 							
	& \times \mathcal{N}(\epsilon_1 k_1,\ldots,\epsilon_N k_N) \sum_P{(-1)^P a(P,\{ \epsilon k \}) e^{i \sum_j \epsilon_{Pj} k_{Pj} x_j}}  \label{prvired} \\
	&=\frac{1}{N!} \int_{k_1>0,\ldots,k_N>0} dk_1\cdots dk_N \ \sum_{\{ \epsilon \}} b(\epsilon_1 k_1,\ldots,\epsilon_N k_N) A'(\epsilon_1 k_1,\ldots,\epsilon_N k_N) \nonumber \\ 	
	&\times \mathcal{N}(\epsilon_1 k_1,\ldots,\epsilon_N k_N) \sum_P{(-1)^P a(P,\{ \epsilon k \}) e^{i \sum_j \epsilon_{Pj} k_{Pj}x_j}} 	\label{drugired} \\
	&=\frac{1}{N!} \int_{-\infty}^{\infty} dk_1 \cdots dk_N \ b(k_1,\ldots,k_N) A'(k_1,\ldots,k_N) \nonumber \\ 
	&\times	\mathcal{N}(k_1,\ldots,k_N) \sum_P{(-1)^P a(P,\{  k \}) e^{i \sum_j k_{Pj}x_j}} \label{trecired}	 \\
	&=\frac{1}{N!} \sum_P \int_{-\infty}^{\infty} dk_1 \cdots dk_N \ b(k_{P1},\ldots,k_{PN}) A'(k_{P1},\ldots,k_{PN}) \nonumber \\
	&\times	 \mathcal{N}(k_{P1},\ldots,k_{PN}) a(P,\{ k \}) e^{i \sum_j k_{Pj}x_j} 	 \label{cetvrtired}\\
	&=\frac{1}{N!} \sum_P \int_{-\infty}^{\infty} dk_{P1} \cdots dk_{PN} \  b(k_{P1},\ldots,k_{PN}) A'(k_{P1},\ldots,k_{PN}) \nonumber \\
	&\times	 \mathcal{N}(k_{P1},\ldots,k_{PN}) a(P,\{  k \}) e^{i \sum_j k_{Pj}x_j}, \label{zadnjired}
\end{align}
\end{widetext}
from which we immediately obtain Eqs. (\ref{psiB0-G}) and (\ref{Gk}) because the sum 
over all permutations $P$ is a sum over $N!$ identical integrals. 
In the derivation above, the first identity, Eq. (\ref{prvired}), follows from the properties
(\ref{perm_G}) and (\ref{b_perm}). The second identity (\ref{drugired}) is due to (\ref{b_eps}) 
and the definition of $A(\{ \epsilon \},\{ k \})$ in Eq. (\ref{Acrtano}).
By employing Eqs. (\ref{minus_G}) and (\ref{b_eps}), 
the sum over $\{\epsilon\}$ in Eq. (\ref{drugired}) can be replaced by 
integrating over the whole $k$-space to obtain the third equality, Eq. (\ref{trecired}). 
By using identities $A'(k_{P1},\ldots,k_{PN})=A'(k_1,\ldots,k_N)$ and
$\mathcal{N} (k_{P1},\ldots,k_{PN})=\mathcal{N} (k_1,\ldots,k_N)$, together with Eq. ($\ref{b_perm}$), 
we obtain ($\ref{cetvrtired}$). 

The time-dependent solution of the many-body Schr\" odinger Eq. (\ref{LLmodel}) 
with $V(x)$ given by (\ref{potential}) is simply 
\begin{equation}
\psi = \int_{-\infty}^{\infty} dk_1 \ldots dk_N G(k_1,\ldots,k_N) e^{i \sum_j (k_j x_j-k_j^2 t)}.
\label{psiB-G}
\end{equation}
Thus, by knowing the function $G$ which contains all information about 
the initial condition, and which is simply related to the projection coefficients 
$b(k_1,\ldots,k_N)$ of the initial state onto hard-wall Lieb-Liniger eigenstates $\phi_{\{ k \}}$,
we can compute the time-dependent Lieb-Liniger wave function in the hard-wall potential 
by employing the Fourier transform. With this identification, an exact analysis 
of this many-body problem is at least conceptually considerably simplified. 

We note that the asymptotic behavior of the many-body state and the 
observables such as single-particle density or momentum distribution can be 
straightforwardly extracted from expression (\ref{psiB-G}) by using the 
stationary phase approximation, as it was done in Refs. \cite{Jukic2008,Jukic2009} 
for the case of free expansion of a Lieb-Liniger gas 
[e.g., see Eq. (15) in Ref. \cite{Jukic2008}, and 
Eqs. (18) and (19) in Ref. \cite{Jukic2009}].
From these methods, and Eqs. (\ref{psiB-G}) and (\ref{Gk}), it follows that the 
initial conditions are imprinted into asymptotic states. 
It is straightforward to infer that the asymptotic wave functions, 
$\psi_{\infty}(\eta_1,\ldots,\eta_N,t)=\psi(\eta_1 t,\ldots,\eta_N t, t)$ 
for sufficiently large $t$, vanish at the hyperplanes of contact between 
particles $\eta_i=\eta_j$ ($i\neq j$), which is characteristic for Tonks-Girardeau 
wave functions \cite{Girardeau1960}. 
However, it should be emphasized that the properties of such asymptotic states 
can considerably differ from the physical properties of a Tonks-Girardeau gas 
in the ground state of some trapping potential \cite{Jukic2008}
(see also the second item of Ref. \cite{Buljan2008}).
Moreover, the asymptotic momentum distribution coincides, up to a simple scaling 
transformation, with the shape of the asymptotic single-particle 
density in $x$-space, reflecting the fact that the dynamics is asymptotically ballistic 
\cite{Jukic2009}; this means that at asymptotic times, despite of the fact that 
the wave functions have attained the Tonks-Girardeau structure, interactions 
do not affect the dynamics any more. 
From the connection between the asymptotic momentum distribution and 
single-particle density one finds that the asymptotic momentum distribution 
is zero at $k=0$, and it is located on the positive $k$-axis, which simply 
means that for sufficiently large times the particles move away from the wall.

%=======================================================================================
\section{Example: A Lieb-Liniger wave packet incident on the hard wall}
\label{sec:example}
%

%
%=========================================================================
\begin{figure}
\begin{center}
\includegraphics[width=0.9 \textwidth]{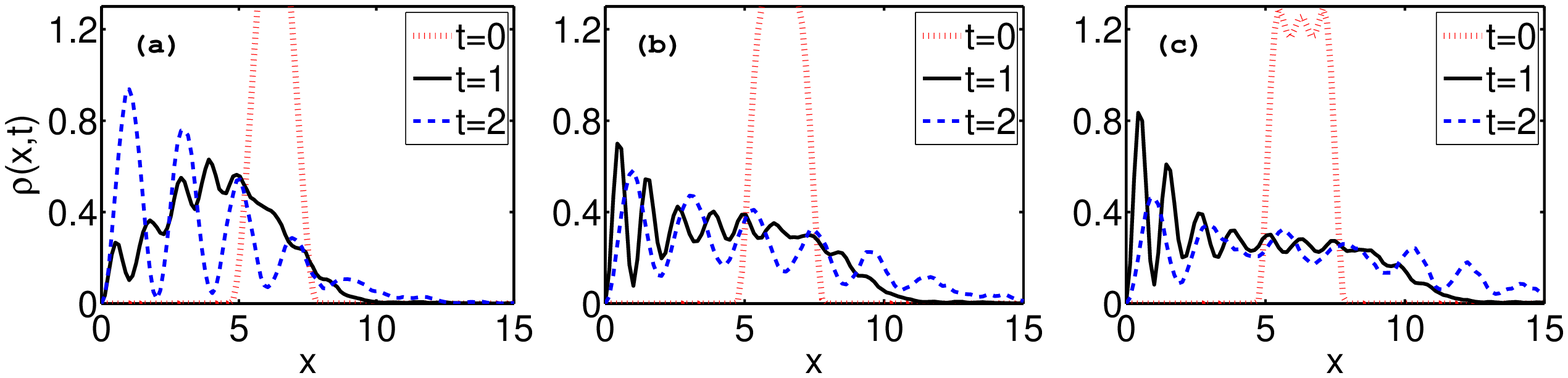}
\caption{ \label{figdensity}
(color online) Density evolution of a Lieb-Liniger wave packet 
comprised of $N=3$ bosons, which is given some momentum kick 
$K$ (per particle) towards the wall. Insets correspond to the 
interaction strengths (a) $c=0.25$, (b) $c=3$, and (c) $c=10$. 
The imparted momentum is $K=1$. Red dotted lines are for $t=0$, 
black solid lines are for $t=1$, and 
blue dashed-lines are for $t=2$.
}
\end{center}
\end{figure}
%=========================================================================
%
%
%=========================================================================
\begin{figure}
\begin{center}
\includegraphics[width=0.9 \textwidth]{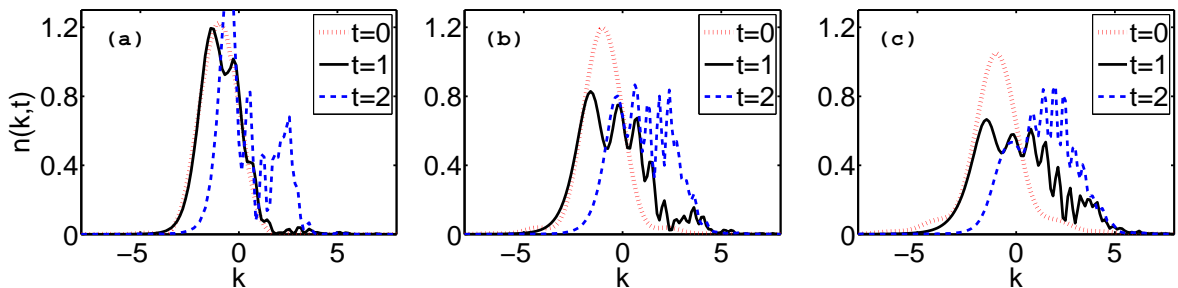}
\caption{ \label{figmomenta}
(color online) The same as in Fig. \ref{figdensity} but for the
momentum distribution}
\end{center}
\end{figure}
%=========================================================================
%

In this section we study a specific example of a localized Lieb-Liniger wave packet 
comprised of a $N=3$ particles reflecting from the hard-wall potential. 
More specifically, we assume that for $t<0$ the Lieb-Liniger system is in the 
ground state of an infinitely deep box denoted by $\psi_{g.s.}(x_1,x_2,x_3)$. 
The analytic expression for this ground state was found in Ref. \cite{Gaudin1971}; 
for reasons of completeness, in Appendix \ref{app:init} we present its construction. 
In our simulations, the box is in the interval $[1.5\pi,2.5\pi]$, i.e., $\psi_{g.s.}(x_1,x_2,x_3)$ is zero 
whenever any $x_i$ is outside of this interval. 
At $t=0$ the box potential is suddenly turned off, and the wave packet 
is simultaneously (and suddenly) imparted some momentum of magnitude $K\geq 0$ towards the wall:
$\psi(x_1,x_2,x_3,t=0)=\psi_{g.s.}(x_1,x_2,x_3) \exp[-iK(x_1+x_2+x_3)]$;
apparently, $K$ denotes the imparted momentum per particle. 
From such an initial state, we are able to find projection coefficients $b(k_1,k_2,k_3)$ 
defined in Eq. (\ref{initB0}), that is, we can find the corresponding function $G(k_1,k_2,k_3)$ 
which is needed to calculate the Fourier transform (\ref{psiB-G}). 
The Fourier integral in (\ref{psiB-G}) is in this particular example $3$-dimensional, 
and it is calculated numerically by using the fast Fourier transform algorithm in MATLAB. 
This provides us with the time-dependent wave function $\psi(x_1,x_2,x_3,t)$, which 
we use to study dynamics of observables such as the single-particle (SP) density $\rho(x,t)$ or 
the momentum distribution $n(k,t)$.

First, let us explore the effect of the interactions on the reflections of 
a few-body Lieb-Liniger wave packet. In Figures \ref{figdensity} and \ref{figmomenta} we plot 
the time-evolution of single-particle densities and distributions of the momenta, respectively. 
The plots are made at three different times, $t=0$, $1$, and $2$, and for three values 
of the coupling parameter, $c=0.25$, $3$, and $10$.
The magnitude of the 
imparted momentum per particle is $K=1$.
Note that the wave packets broaden in time 
due to the repulsive interactions between the particles, and also due to the 
wave dispersion effects; the wave packets for larger values of $c$ 
spread at a faster rate than the wave packets for smaller $c$. 
From Figs. \ref{figdensity} and \ref{figmomenta} we observe that wave packets 
with a larger interaction parameter $c$ get reflected faster than the wave packets 
for smaller $c$; for wave packets with smaller repulsion between the particles 
(smaller $c$), the compression of the wave packet is stronger, and therefore 
reflection of the momenta occurs at a slower rate. 
We also observe that all wave packets exhibit interference fringes 
during the reflection process. However, we find the interference fringes 
to be deeper for smaller values of $c$, which follows from the 
fact that the wave packets for smaller $c$ are more spatially coherent. 
This can be seen also from Fig. \ref{figmomenta} which displays momentum distributions. 
The distribution $n(k,t)$ for $c=0.25$, at the largest time shown $t=2$,  
has one strong well-defined peak (the one closest to zero), and several smaller 
peaks of the wave components with larger magnitude of the momentum [see 
Fig. \ref{figmomenta}(a)]. 
In contrast, for $c=10$ this most dominant peak close to $k=0$ is much smaller
[see Fig. \ref{figmomenta}(c)]. 

%
%=========================================================================
\begin{figure}
\begin{center}
\includegraphics[width=0.9 \textwidth]{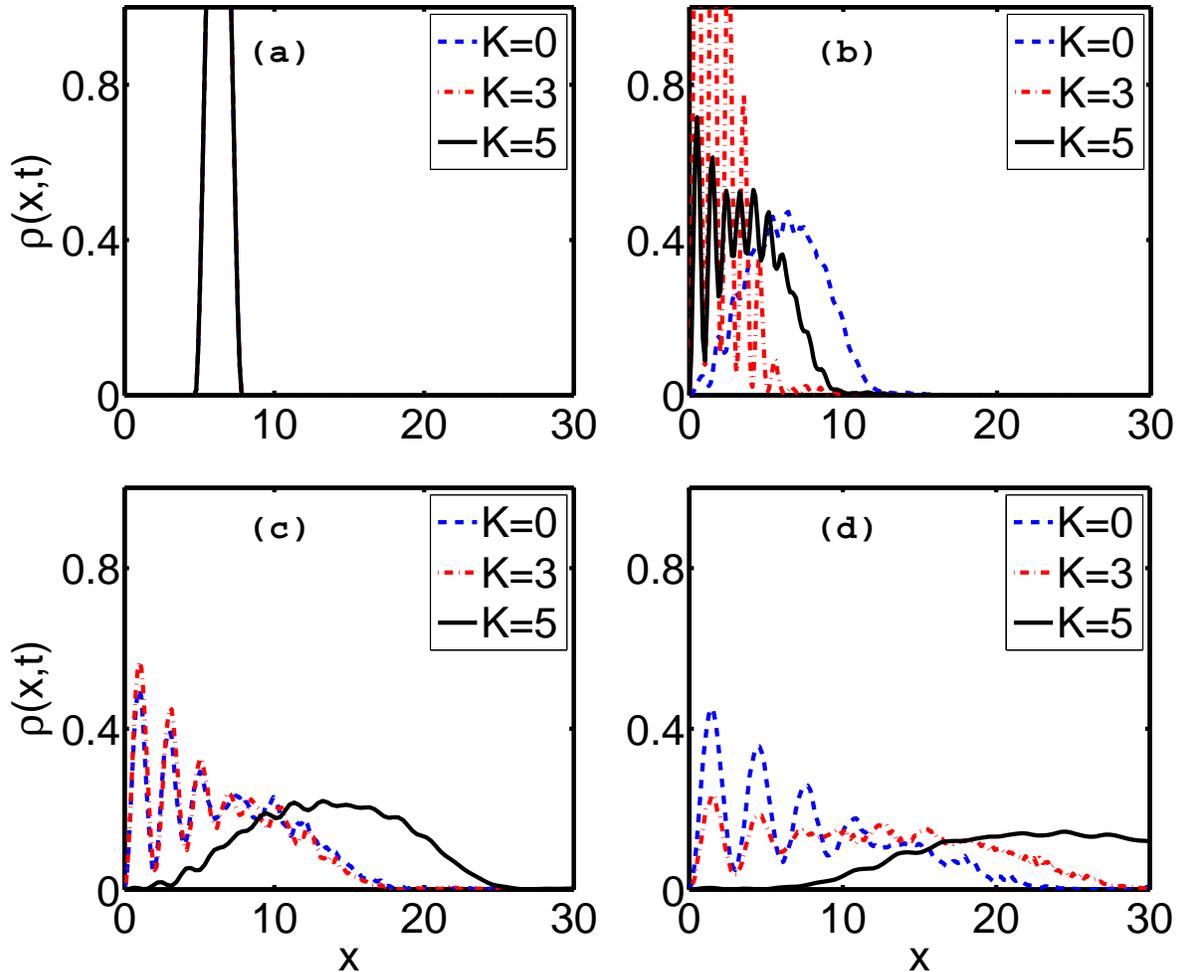}
\caption{ \label{figdensity1}
(color online) Density evolution of a Lieb-Liniger wave packet 
comprised of $N=3$ bosons, which is given some momentum kick 
$K$ (per particle) towards the wall. Insets correspond to the 
times (a) $t=0$, (b) $t=1$, (c) $t=2$, and (d) $t=3$. 
The interaction strength is $c=1$. Blue dashed-lines are for $K=0$, 
red dot-dashed lines are for $K=3$, and black solid lines are for $K=5$.}
\end{center}
\end{figure}
%=========================================================================
%
%
%=========================================================================
\begin{figure}
\begin{center}
\includegraphics[width=0.9 \textwidth]{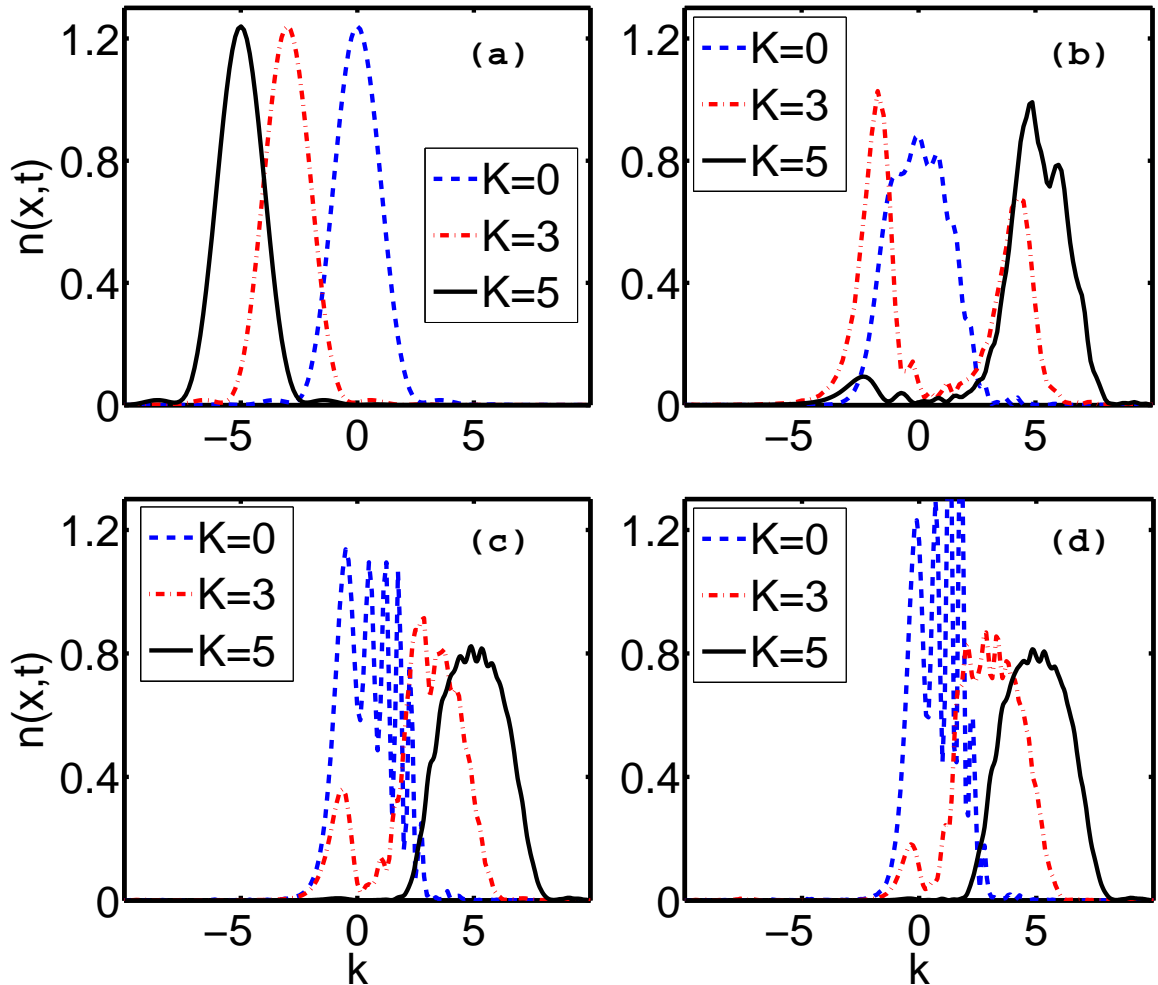}
\caption{ \label{figmomenta1}
(color online) The same as in Fig. \ref{figdensity1} but for the
momentum distribution} 
\end{center}
\end{figure}
%=========================================================================
%

Next we explore dependence of the time-evolution on the imparted momentum. 
To this end we fix the interaction strength at $c=1$, and observe the time-evolution 
for three different initial conditions (see Figs. \ref{figdensity1} and \ref{figmomenta1}): 
(i) expansion in the presence of the wall occurs when $K=0$, 
(ii) reflection at an intermediate value $K=3$, and (iii) for 
large value of the imparted momentum $K=5$. 
The wave packets for $K=3$ and $5$ have the property that basically all of the 
initial momentum distribution is directed towards the wall, i.e., the distributions 
at $t=0$ is on the negative $k$-axis. In contrast, exactly half of the initial momentum 
distribution of the wave packet for $K=0$ is positive (negative). 
The basic distinction between these cases is that the wave packets with 
sufficiently large imparted momentum $K$ get simply reflected from the 
wall and at larger times the interference fringes are almost negligible. 
For example, the wave packet with $K=5$ is practically completely reflected 
from the wall at $t=2$, see solid black lines in Figs. \ref{figdensity1}(c) and \ref{figmomenta1}(c); 
the momentum distribution is on the positive $k$-axis and the interference 
fringes are essentially absent. 
In contrast, for $K=0$ half of the momentum distribution is already 
positive (corresponding to motion away from the wall), and this part 
interferes with the reflected component at all times of the evolution. 
Note that the wave packet with $K=0$ is still in the process of reflection from the 
wall at $t=2$ because a large fraction of its momentum distribution is 
still on the negative $k$-axis, see dashed blue line in Fig. \ref{figmomenta1}(c); 
the interference fringes are the largest in this case, 
see dashed blue line in Fig. \ref{figdensity1}(c).

%
%=========================================================================
\begin{figure}
\begin{center}
\includegraphics[width=0.9 \textwidth]{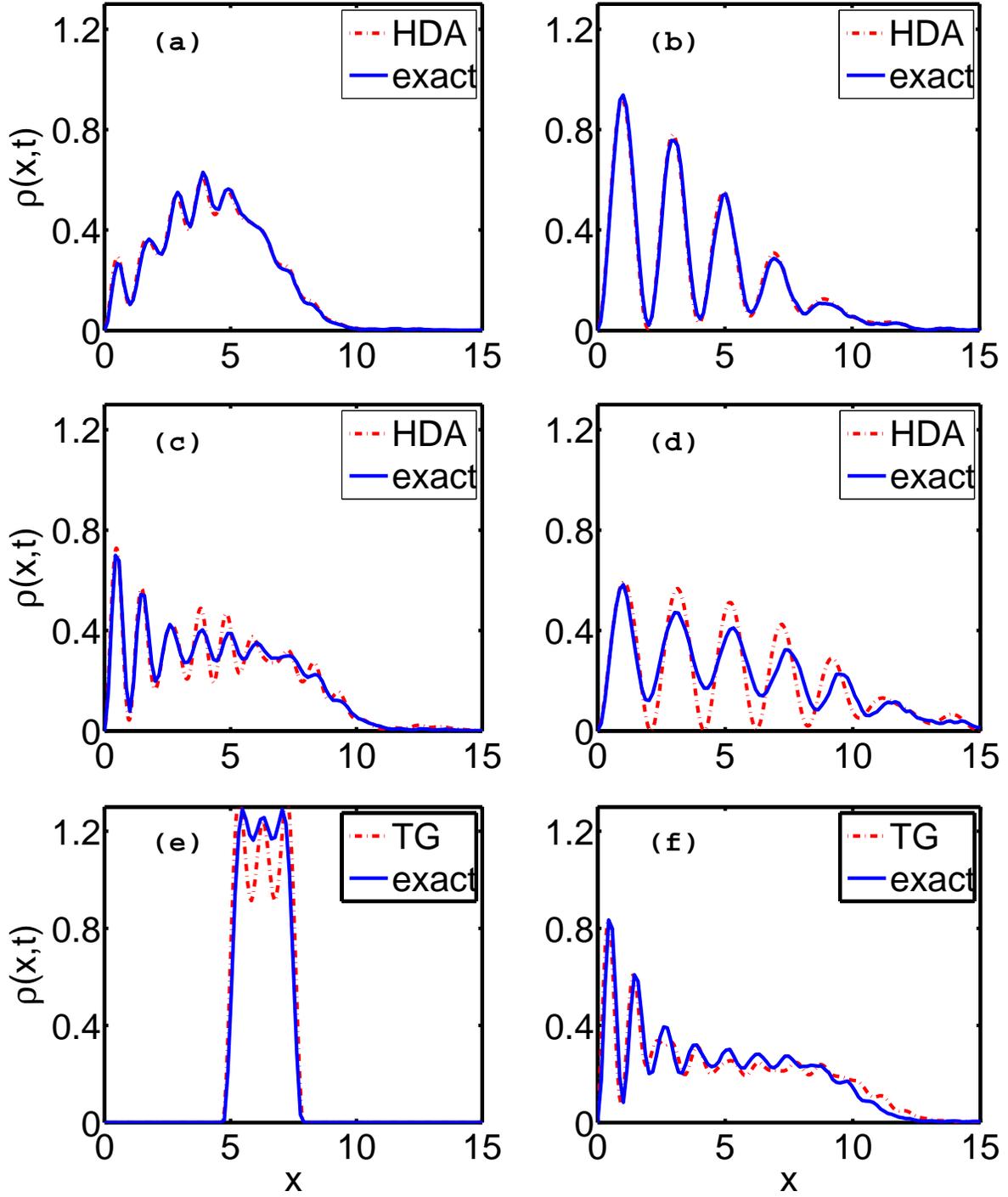}
\caption{ \label{figcomp}
(color online) Comparison of the density evolution in the 
exact calculation, with the hydrodynamic approximation (HDA) 
[(a)-(d)], and the Fermi-Bose mapping [(e) and (f)] valid in the 
Tonks-Girardeau (TG) regime. 
The interaction strengths $c$ and times $t$ in the insets are: 
(a) $c=0.25$, $t=1$; 
(b) $c=0.25$, $t=2$;
(c) $c=3$, $t=1$; 
(d) $c=3$, $t=2$;
(c) $c=10$, $t=0$; 
(d) $c=10$, $t=1$.
The initially imparted momentum is $K=1$ for all figures.} 
\end{center}
\end{figure}
%=========================================================================
%

Exact solutions can serve as a benchmark to check the range of validity of 
other methods which may be used to analyze nonequilibrium dynamics of 
interacting systems. We have compared the solutions obtained with the 
Fourier transform method presented here with the so-called hydrodynamic formalism
\cite{Dunjko}, which describes the Lieb-Liniger system via the nonlinear 
Schr\"odinger equation with variable nonlinearity \cite{Ohberg2002}.
In Figs. \ref{figcomp} (a)-(d), we show density profiles for two different couplings 
($c=0.25$ and $c=3$) at two different times ($t=1$ and $t=2$).
We find that the single-particle density (and momentum distribution), calculated
within this method, are in good agreement with our simulations for small values 
of the coupling parameter $c$ (up to $c=1$); this upper limit for $c$ also depends 
on the initial density of the 1D Bose gas, as it is well known that the effective 
interaction strength parameter is $c$ divided by the linear density \cite{Lieb1963}. 
However, for larger values of $c$, the hydrodynamic formalism goes beyond its range 
of validity for the simulations presented here.
For example, for the simulations at intermediate interaction strength $c=3$ 
[see Figs. \ref{figcomp} (c) and (d)], the hydrodynamic formalism predicts deeper 
interference fringes than those obtained via the Fourier transform method; this 
is attributed to the fact that the hydrodynamic formalism overestimates the spatial 
coherence of the wave packet \cite{Dunjko,Ohberg2002}. 
For sufficiently large $c$, the system is in the Tonks-Girardeau regime, and one can employ 
the Fermi-Bose mapping \cite{Girardeau1960,Girardeau2000} to study the dynamics.
In Fig. \ref{figcomp} (e) and (f) we compare our calculation with that obtained 
via Fermi-Bose mapping ($c=\infty$, \cite{Girardeau1960,Girardeau2000}) for a 
large value of the interaction strength $c=10$; we observe that qualitative features 
of the Tonks-Girardeau regime such as the $N$ peaks in the initial single-particle 
density coincide in the two calculations, however, even larger $c$ is needed 
to obtain better quantitative agreement. 

\subsection{Normalization of eigenstates}

In order to numerically check our conjecture that the Lieb-Liniger hard-wall 
eigenstates defined in (\ref{psiHW}) are properly normalized, we have compared 
the initial state obtained via $\psi_{g.s.}(x_1,x_2,x_3)\exp[-iK(x_1+x_2+x_3)]$, and 
the wave function obtained via Eq. (\ref{psiB0-G}) by employing the function $G(k_1,k_2,k_3)$, 
which is calculated from the projection coefficients $b(k_1,k_2,k_3)$ as in Eq. (\ref{Gk}). 
We found that the relative agreement between the two wave functions is on the order of 
$1$\% or better, which is on the order of the numerical accuracy for the size of 
our numerical grid, which is limited by computer memory. We have performed this 
comparison for various initial conditions (different $K$ and $c$ values). Unfortunately, 
a rigorous proof of normalization of Lieb-Liniger hard-wall eigenstates is 
to the best of our knowledge still lacking. 

\subsection{Connection to physical units}

If we consider a system of $^{87}$Rb atoms, then the ratio 
$X_0^2/T_0 \approx 3.65 \times 10^{-10}$m$^{2}$/s is fixed. 
By choosing for example $X_0 \approx 1.35\, \mu$m for the spatial scale, the 
temporal scale is set to $T_0=5$ms. The 3D scattering length is $a=5.3$nm. 
The interaction parameter $c$ can be varied by changing the width of transversal 
confinement $l_{\perp}$; for example, the values of $c=0.25$ up to $c=10$, can 
be obtained by varying $l_{\perp}$ from $242$nm down to $l_{\perp}\approx41$nm, respectively.
Of course, for a different choice of temporal and spatial scales, transversal 
confinements $l_{\perp}$ would have different values. We have also verified that for 
the choice of scales in our example, the longitudinal energy $E_0$ is less then 
the transverse energy spacing $\hbar \omega_{\perp}$, a condition needed for freezing 
the radial degrees of freedom.

%=======================================================================================
\section{Conclusion}
\label{sec:concl}

We have studied reflections of a Lieb-Liniger wave packet from the hard-wall potential. 
By employing the symmetry of the many-body eigenstates with respect to the 
change of the sign and permutation of their quantum numbers (i.e., quasimomenta), 
that is, Equations (\ref{perm_G}) and (\ref{minus_G}), we have demonstrated that 
time-evolution of this interacting many-body wave packet can be represented in terms 
of an $N$-dimensional Fourier transform, where $N$ is the number of particles in the 
wave packet. This result simplifies our understanding of the time-evolution in 
this many-body problem and enables straightforward calculation of the time-asymptotic 
properties of the system. 

We have utilized the formalism to numerically study dynamics of single-particle density and 
momentum distribution of a few-body wave packet reflecting from the wall 
(the wave packet is initially close to the wall). 
Reflection dynamics and interference phenomena depend on the strength of the interaction 
between the particles $c$ and the imparted momentum $K$ towards the wall. 
The wave packets for smaller $c$ get reflected at a slower rate, because 
they get compressed more strongly as the wave packet hits the wall. 
Moreover, the interference fringes are deeper (larger visibility) for smaller 
values of $c$. 
If $K$ is sufficiently large such that the initial momentum distribution 
is on the negative $k$-axis, the wave packet gets reflected and the interference 
fringes become small as soon as most of the momenta become positive. 
On the other hand, for $K=0$, the interference effects are fairly large.

%=======================================================================================

\acknowledgments
This work is supported by the Croatian Ministry of Science (Grant No. 119-0000000-1015), 
the Croatian National Foundation for Science and the Croatian-Israeli project cooperation. 
We are grateful to Adolfo del Campo and Ofir Alon for useful comments and suggestions.

%=======================================================================================
\begin{appendix}
\section{The ground state of a Lieb-Liniger gas in an infinitely deep box \cite{Gaudin1971}}
\label{app:init}

In section \ref{sec:example} we study Lieb-Liniger dynamics in the presence of the 
hard wall potential, with an example of three Lieb-Liniger bosons which are at $t<0$ confined
in the ground state of an infinitely deep box of length $L=\pi$. 
The ground state in fundamental permutation sector $R_1$
has been constructed by Gaudin in Ref. \cite{Gaudin1971} via a
superposition of $2^N$ free space eigenstates. For the box in 
the interval $[1.5 \pi,2.5 \pi]$, the ground state (up to a normalization constant) reads
\begin{equation}
\psi_{g.s.}(x_1,\ldots,x_N)\propto \sum_{\{\epsilon\}} \epsilon_1 \cdots \epsilon_N \prod_{i<j} \left( 1- \frac{ic}{q_i+q_j} \right)
		\sum_{P}  \prod_{i<j} \left( 1+ \frac{ic}{q_{Pi}-q_{Pj}} \right) e^{i \sum_j q_{Pj} (x_j-1.5 \pi)}.
\end{equation}
Here, summations are taken over $2^N$ elements of set $\{\epsilon\}$, and $N!$ permutations $P$.
The quasimomenta $q_j= \epsilon_j |q_j|$, for $j=1,\ldots,N$, are determined by set
of transcendental equations
\begin{equation}
|q_i| L = \pi + \sum_{j \neq i} \left( \tan^{-1} \frac{c}{|q_i|-|q_j|} + 
\tan^{-1} \frac{c}{|q_i|+|q_j|}\right).
\label{transc}
\end{equation}
Eqs. (\ref{transc}) are solved numerically. For the initial state corresponding 
to three particles ($N=3$), where $\psi(x_1,x_2,x_3,t=0)=\psi_{g.s.}(x_1,x_2,x_3) \exp[-iK(x_1+x_2+x_3)]$, 
it is straightforward to obtain the projection coefficients $b(\{k\})$ by 
employing Eq. (\ref{initB0}) and the orthonormality of eigenstates $\phi_{\{k\}}$. 

\end{appendix}

%==============================================================================

\end{document}